\documentclass[%
 reprint,
nofootinbib,
 amsmath,amssymb,
 aps,
]{revtex4-2}

\usepackage{graphicx}
\usepackage{dcolumn}
\usepackage{bm}

\usepackage{subfigure,graphicx,epsfig,amsmath,amsfonts,amssymb,xcolor,slashed,ulem,multirow}

\usepackage{relsize}
\usepackage{slashed}
\usepackage{color}
\usepackage{ulem}
\usepackage{tabu}

\usepackage{amsmath}
\usepackage{amssymb}
\usepackage{amsthm}
\usepackage{mathrsfs}
\usepackage{graphicx}
\usepackage{epstopdf}
\usepackage{fancyhdr}
\usepackage{array}
\usepackage[all]{xy}
\usepackage{eufrak}
\usepackage{euscript}
\usepackage{enumerate}
\usepackage{slashed}
\usepackage{hyperref}
\usepackage{subfigure} 
\usepackage{epstopdf} 

\usepackage{hyperref}
\hypersetup{pdftex,colorlinks=true,linkcolor=blue,citecolor=blue,menucolor=black,urlcolor=blue,filecolor=blue}

\newcommand{\be}{\begin{equation}}
\newcommand{\ee}{\end{equation}}
\newcommand{\ba}{\begin{eqnarray}}
\newcommand{\ea}{\end{eqnarray}}
\newcommand{\nn}{\nonumber}

\begin{document}


\title{Study of hidden-charm, doubly-strange pentaquarks in $\Lambda_b\to J/\psi \Xi^- K^+$ and $\Xi_b\to J/\psi \Xi^- \pi^+$}

\date{\today}

\author{L.~Roca}
\email[]{luisroca@um.es}
\affiliation{Departamento de F\'isica, Universidad de Murcia, E-30100 Murcia, Spain}

\author{J.~Song}
\email{song-jing@buaa.edu.cn}
\affiliation{Center for Theoretical Physics, School of
 Physics
 and Optoelectronic 
 Engineering, Hainan
University, Haikou 570228, China}

\author{E.~Oset}
\email{oset@ific.uv.es}
\affiliation{Departamento de F\'{\i}sica Te\'orica and IFIC, Centro Mixto Universidad de
Valencia-CSIC Institutos de Investigaci\'on de Paterna, Aptdo.22085,
46071 Valencia, Spain}
\affiliation{Guangxi Key Laboratory of Nuclear Physics and Technology, Guangxi Normal University, Guilin 541004, China}

\begin{abstract}

Hidden-charm pentaquark states with double strangeness, $P_{css}$, have been predicted within the framework of unitary coupled-channel dynamics. In this work, we theoretically investigate the potential to observe these states in the decays $\Lambda_b\to J/\psi \Xi^-K^+$ and $\Xi_b\to J/\psi\Xi^-\pi^+$. In this framework, these pentaquark configurations couple strongly to the $J/\psi\Xi$ channel, as well as to other vector-baryon channels with the $\bar c c s s n$ flavor structure, making these decay modes promising for their observation through the corresponding invariant-mass distributions. Our analysis begins with the identification of the dominant weak decay mechanisms, followed by hadronization into meson-baryon channels, connected through flavor symmetry. Final-state interactions are then incorporated to dynamically generate the full amplitude, accounting for the formation of the pentaquark states. We compare our results with recent LHCb measurements of the $J/\psi\Xi^-$ mass distribution and find that, given the predicted pentaquark width of about 10 MeV in this channel, the state is too narrow to be resolved with the current experimental resolution, but it would become visible with significantly improved mass precision.

\end{abstract}

\maketitle


\section{Introduction}

The spectroscopy of heavy-quark pentaquark states has undergone remarkable progress in recent years, driven in part by the LHCb discoveries of hidden-charm pentaquarks such as $P_c(4312)$, $P_c(4440)$, $P_c(4457)$, and their strange partners $P_{cs}(4459)$, and $P_{cs}(4338)$~\cite{LHCb:2015yax,LHCb:2019kea,LHCb:2020jpq,LHCb:2022ogu,LHCb:2021chn}.
Some of these observations confirmed long-standing predictions that meson-baryon interactions could generate molecular-like configurations beyond the conventional three-quark picture~\cite{Wu:2010jy,Wu:2010vk,Wu:2012md,Xiao:2013yca,Karliner:2015ina}.
The experimental findings triggered extensive theoretical efforts to understand their nature, with a particular emphasis on hadronic molecular configurations (see reviews in \cite{Chen:2016qju,Lebed:2016hpi,Guo:2017jvc,Liu:2019zoy,Ali:2017jda,Chen:2022asf}).

An unexplored frontier is the possible existence of doubly-strange hidden-charm partners, usually denoted $P_{css}$, with flavor content $\bar ccssn$.
Although no experimental signal has been reported so far, several  vector-meson coupled-channel unitary models  predict the existence of such states in the 4.5-4.7~GeV region \cite{Marse-Valera:2022khy,Roca:2024nsi}. Complementary studies have been carried out within alternative frameworks, including meson-exchange models \cite{Wang:2020bjt}, sum rules \cite{Azizi:2021pbh,Ozdem:2022iqk,Ozdem:2023htj}, and quark models \cite{Anisovich:2015zqa,Ortega:2022uyu,Germani:2024miu,Liu:2025slt}.

Theoretical predictions of such $P_{css}$ pentaquarks naturally motivate experimental searches, as exemplified by the earlier discovery of the single strange partner $P_{cs}$. This state had been anticipated within coupled-channel unitary approaches \cite{Wu:2010jy,Wu:2010vk,Xiao:2019gjd}, and the decay $\Xi_b^-\to J/\psi K^-\Lambda$ was proposed in Ref.~\cite{Chen:2015sxa} as a promising production channel. The LHCb collaboration later confirmed this prediction, observing $P_{cs}$ in the corresponding invariant-mass spectrum for the first time \cite{LHCb:2020jpq}.

The lowest $P_{css}$ state predicted  in  Ref.~\cite{Roca:2024nsi} arises from the 
pseudoscalar-baryon ($PB$) interaction with quantum numbers $I(J^P)=\frac{1}{2}(\frac{1}{2}^-)$ and flavor content  $\bar c c s s n$,
which was associated to a molecular pentaquark state with a mass around 4500~MeV and a width of order 10~MeV, coupling mainly  to $\bar D \Omega_c$ and $\bar D_s\Xi'_c$, and more weakly to $\eta_c \Xi $. 
Ref.~\cite{Oset:2024fbk} further showed that such a state could manifest in the decays $\Xi_b^0\to\eta\eta_c\Xi^0$ and $\Omega_b^-\to K^-\eta_c\Xi^0$,  though a realistic observation would likely require the higher luminosity foreseen in the LHCb Upgrade~2 project \cite{futureLHCb}.

Building on this line of investigation, it is natural to explore other decays where $P_{css}$ states might appear. At this point, we can take advantage of the recent LHCb experimental results on the decays $\Lambda_b\to J/\psi \Xi^- K^+$ and $\Xi_b\to J/\psi \Xi^- \pi^+$ \cite{LHCb:2025lhk}, which provides $J/\psi \Xi^-$ invariant-mass spectra that could reveal the presence of $\bar c c s s n$ resonances coupling to this channel.
Indeed, Ref.~\cite{Roca:2024nsi} predicts an additional $P_{css}$ resonance in the $4.6$-$4.7$~GeV region with a width of about $10$~MeV, dynamically generated in the vector-baryon ($VB$) coupled channels $J/\psi\Xi$, $\bar D^*_s \Xi'_c$, and $\bar D^* \Omega_c$, with $J^P=1/2^-$, $3/2^-$. If present, this resonance should leave a signature in the $J/\psi\Xi^-$ spectrum of the aforementioned decays. 
In Ref.~\cite{Roca:2024nsi}
 the two spin states appear degenerate. In practice, this degeneracy could be broken and two near peaks could be observed, in analogy with the two $P_c$ states, $P_c(4440)$ and $P_c(4457)$ which are associated to the two possible spins $3/2^-$ and $1/2^-$ as a $\bar D^* \Sigma_c$ molecular state \cite{Liu:2019tjn,Guo:2019kdc,Du:2021fmf,Du:2019pij} (see a recent update on the issue in the work of \cite{Yang:2024nss}).

In this work we perform a theoretical analysis of the $J/\psi\Xi^-$ mass distribution in $\Lambda_b\to J/\psi\Xi^-K^+$ and $\Xi_b\to J/\psi\Xi^-\pi^+$.
We first identify the dominant weak decay mechanisms at the quark level, followed by hadronization through the creation of $q\bar q$ pairs in the spirit of the $^3P_0$ model. The resulting meson-baryon channels are related via flavor symmetry at the quark level, and the final-state vector-baryon interaction is incorporated to dynamically generate the required $P_{css}$ state.
We show the predicted $J/\psi\Xi^-$ mass spectra and compare them with the experimental distributions after convolution with the LHCb bin resolution. Given the current experimental resolution, exceeding $100$ and $200$ MeV, the narrow predicted resonance ($\sim 10$ MeV) cannot be distinguished. We thus discuss the resolution requirements and the expected signatures that could allow such states to be identified in future higher precision measurements.

\section{Formalism}

\subsection{Overview of the unitary coupled channel model}

We begin with a brief summary of the framework used to generate the $P_{css}$ states, following Ref.~\cite{Roca:2024nsi}, where full details can be found. The approach is based on the chiral unitary method \cite{Kaiser:1995eg,Oset:1997it,Oller:1998zr,Oller:2000fj,Dobado:1996ps,Oller:1998hw,Nieves:2001wt}, which implements unitarity in coupled channels and has been successfully applied to a broad spectrum of hadronic resonances in the past 25 years. In the present case, the relevant channels are $J/\psi\Xi$, $\bar D^*_s \Xi'_c$, and $\bar D^* \Omega_c$, which are coupled through interaction kernels derived from $t$-channel vector-meson exchange.  

The required vertices are constructed within the local hidden-gauge formalism \cite{Bando:1984ej,Bando:1987br,Birse:1996hd,Meissner:1987ge,Nagahiro:2008cv}, consistently extended to the charm sector \cite{Molina:2009ct,Molina:2010tx}, and supplemented by constraints from heavy-quark spin symmetry. The scattering amplitudes are obtained by solving the on-shell Bethe-Salpeter equation, 
\begin{equation} t=(1-VG)^{-1}V \ ,
\label{eq:BS}
\end{equation}
where the matrix $V$ accounts for the interaction kernels and $G$ for the meson-baryon loop functions.

The S-wave vector-baryon ($VB$) interaction potentials at tree level, $V_{ij}$, are given by \cite{Roca:2024nsi}
\begin{eqnarray}
  \label{eq:def_Vij}
     V_{ij}=g^2 C_{ij}(p^0+p'^{0}) \, ,
\end{eqnarray}
where $i,j$ label the interacting $VB$ channels and $p^0$ ($p'^0$) denotes the on-shell energy of the incoming (outgoing) vector meson in the center-of-mass frame. The coupling constant $g$ is defined in Ref.~\cite{Roca:2024nsi}.
For channels involving external vector mesons, a global $\vec{\epsilon}_i \cdot \vec{\epsilon}_j$ term has been factored out, and contributions of order $\mathcal{O}(q^2/M_V^2)$ in the internal vector-meson propagators have been neglected.
The symmetric coefficients $C_{ij}=C_{ji}$ in Eq.~\eqref{eq:def_Vij}, which determine the interaction strengths between the different channels, are listed in Table~\ref{Tab:CijPBVPu}.
\begin{table}[h!]
  \begin{center}
    \begin{tabular}{c|cccc}
        \hline\\ [-0.30cm]
         & $J/\psi\Xi$ &  $\bar{D^*}_s\Xi_c^\prime$ & $\bar{D^*}\Omega_c$\\
        \hline \\ [-0.2cm]
        $J/\psi\Xi$ & $0$ & $\frac{1}{\sqrt{6}m^2_{D^*_s}}$ & $-\frac{1}{\sqrt{3} m^2_{D^*}}$ \\ 
        $\bar{D^*}_s\Xi_c^\prime$ & &$\frac{1}{m^2_\phi}-\frac{1}{m^2_{J/\psi}}$ & $\frac{\sqrt{2}}{m^2_{K^*}}$ \\
        $\bar{D^*}\Omega_c$ & & & $-\frac{1}{m^2_{J/\psi}}$ \\
    \end{tabular}
  \end{center}
\caption{$C_{ij}$ coefficients of the $VB$ interaction in the $\bar c c s s n$ sector.}    
\label{Tab:CijPBVPu}
\end{table}
It is worth noting  that the interaction strengths are inversely proportional to the squared masses of the vector mesons exchanged in the $t$-channel, reflecting the suppression of heavy-meson exchange.

The vector-baryon loop function $G_l$ in Eq.~\eqref{eq:BS} is regularized using a three-momentum cutoff $\Lambda$,
\begin{eqnarray}
        G_l(\sqrt{s})  =  \int_{q< \Lambda} &&\frac{d^3\vec q}{(2\pi)^3}\frac{1}{2\omega_l({q})}\frac{M_l}{E_l({q})} \nonumber\\
                &\cdot&  \frac{1}{\sqrt{s}-\omega_l({ q}) - E_l(q)+i\epsilon}\,,
          \label{eq_Gl}
\end{eqnarray}	  
where $q = |\vec q|$, $\omega_l(q) = \sqrt{m_l^2 + q^2}$, $E_l(q) = \sqrt{M_l^2 + q^2}$, and $m_l$ ($M_l$) are the meson (baryon) masses.
The cutoff $\Lambda$ is the main source of model uncertainty and is taken in the range $600$-$800$ MeV in Ref.~\cite{Roca:2024nsi}.

As discussed in Ref.~\cite{Oset:2024fbk}, for heavy-hadron loops, the cutoff method can make the real part of $G_l$ unreliable near $\sqrt{s} \sim \sqrt{\Lambda^2 + m_l^2} + \sqrt{\Lambda^2 + M_l^2}$ \cite{Wu:2010rv,Xiao:2013jla,Feijoo:2022rxf} where an unphysical pole appears. 
However, since the imaginary part dominates in this region, it suffices to adopt a similar approach to that of \cite{Dai:2020yfu,Oset:2024fbk}, setting the real part to zero whenever it becomes positive.

For $I(J^P)=\tfrac{1}{2}(\tfrac{1}{2}^-,\tfrac{3}{2}^-)$ in the $VB$ channels,  Ref.\cite{Roca:2024nsi} reported a dynamically generated pentaquark with mass and width depending on the cutoff: for $\Lambda=600$ MeV, $M = 4675$ MeV and $\Gamma = 10$~MeV, while for $\Lambda=800$ MeV, $M = 4617$~MeV and $\Gamma = 12$~MeV. The differences between these values provide an estimate of the model uncertainty. The resonance couples most strongly to $\bar D^* \Omega_c$ and $\bar D^*_s \Xi'_c$, with a weaker coupling to $J/\psi \Xi$, which is the only open channel at the resonance energy and thus dominates the decay width. These properties arise from the nonlinear dynamics implicit in Eq.\eqref{eq:BS}, which generate poles in the second Riemann sheet of the scattering amplitudes. In this work, we focus on the production of this pentaquark in the decays $\Lambda_b\to J/\psi \Xi^- K^+$ and $\Xi_b\to J/\psi \Xi^- \pi^+$.

\subsection{Decays $\Lambda_b\to J/\psi \Xi^- K^+$ and $\Xi_b\to J/\psi \Xi^- \pi^+$}

We first present the formalism for the $\Lambda_b \to J/\psi \Xi^- K^+$ decay, while the analogous $\Xi_b \to J/\psi \Xi^- \pi^+$ process will be addressed later.

\begin{figure}[h]
\centering
\includegraphics[width=.95\linewidth]{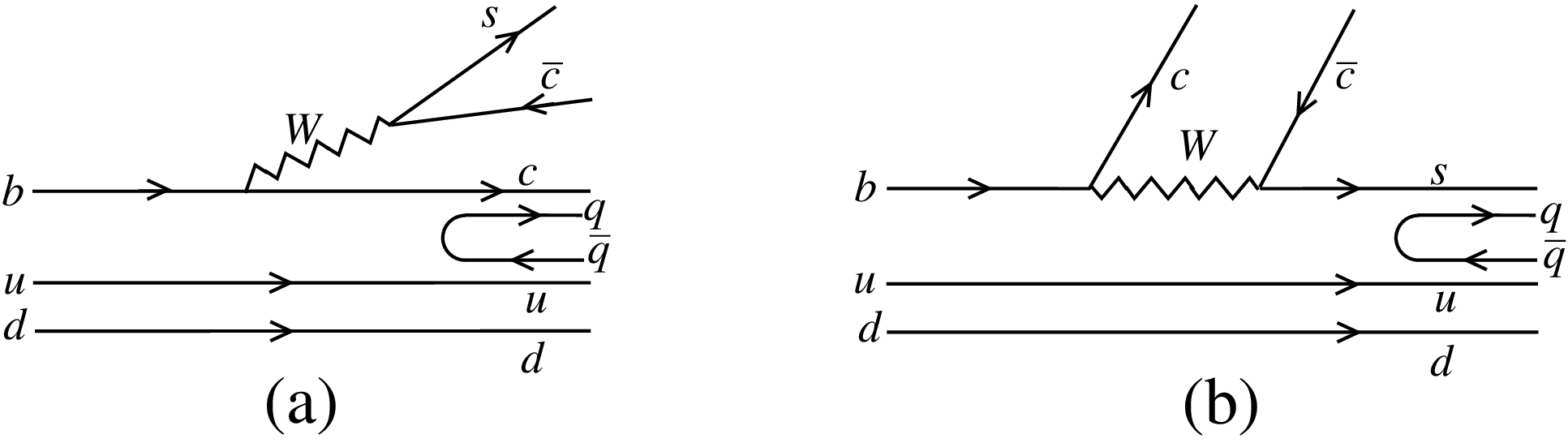}
\caption{Dominant quark-level diagrams for $\Lambda_b\to J/\psi MB$ prior to final-state interactions: (a) external W-emission, (b) internal W-emission.}
\label{fig:diagram_tree_Lambda}
\end{figure}

In line with earlier analyses of related decays, such as $\Xi_b^-\to J/\psi K^- \Lambda$ \cite{Chen:2015sxa}, $\Lambda_c^+\to\pi^+ MB$ \cite{Miyahara:2015cja}, $\Lambda_b\to J/\psi \Lambda(1405)$ \cite{Roca:2015tea}, and more recently $\Xi_b^0\to \eta \eta_c \Xi^0$ and $\Omega_b^-\to K^- \eta_c \Xi^0$ \cite{Oset:2024fbk}, all following the methodology first outlined in \cite{MartinezTorres:2009uk}, the tree-level production at the quark level is driven mainly by the two mechanisms shown in Fig.~\ref{fig:diagram_tree_Lambda}.

Diagram (a) corresponds to external W-emission, while (b) depicts the internal W-emission topology.

Let us discuss first the external W-emission process. In a first step the weak transition converts the $b$ quark of the decaying $\Lambda_b$ into a $c$ quark, and the $W^-$ couples to $s\bar c$.
The resulting $s\bar c $ pair combines to form a $D^{*-}_s$ meson, while the remaining virtual $cud$ configuration hadronizes into a meson-baryon pair. Hadronization is implemented by inserting a $q\bar q$ pair with the quantum numbers of the vacuum, following the philosophy of the $^3P_0$ model \cite{Micu:1968mk,LeYaouanc:1972vsx}.

To evaluate the corresponding amplitude, we start by considering the flavor-spin wave function of the initial $\Lambda_b^0$ baryon  \cite{Capstick:1986ter,Roberts:2007ni,Wang:2022aga}:

\begin{align*}
|\Lambda_b^0\rangle=\frac{1}{\sqrt{2}}|b(ud-du)\rangle\,\chi_{MA}(23),
\end{align*}
where $\chi_{MA}(23)$ is the mixed-antisymmetric spin wave function under exchange of quarks 2 and 3 \cite{close_book,Wang:2022aga}. Consequently, 
the light $u$ and $d$ quarks are antisymmetric in both flavour and spin.
After the weak transition, but prior to hadronization, the three-quark flavor state becomes
\begin{align*}
\frac{1}{\sqrt{2}}|c(ud-du)\rangle\,\chi_{MA}(23),
\end{align*}
preserving the antisymmetric flavor and spin correlation, since the $u$ and $d$ quarks act as spectators.

Hadronization is then implemented by inserting a $q\bar q$ pair with vacuum quantum numbers, leading to the quark-level state in flavour space
\begin{align*}
|H\rangle&\equiv \frac{1}{\sqrt{2}} D^{*-}_s|c\,(u\bar u +d\bar d +s\bar s +c\bar c)\,(ud-du)\rangle\\
         &=\frac{1}{\sqrt{2}} D^{*-}_s 
	|cu(\bar u u d-\bar u d u ) + cd(\bar d u d-\bar d d u) \nn\\
	&+ cs(\bar s u d- \bar s d u) + cc(\bar c u d- \bar c d u)	 
	 \rangle\,,
\end{align*}
where the upper $c$ quark is assumed to form a baryon together with the newly created $q$ and one of the spectator quarks, while the corresponding antiquark combines with the remaining quark to form a meson.

We introduce the quark field vector and the corresponding $q\bar q$ matrix,
\begin{equation}
q\equiv \left(\begin{array}{c}u\\d\\s\\c\end{array}\right)\,\text{~~and~~~}
P\equiv q\bar q^\tau=\left(\begin{array}{cccc}
u\bar u & u\bar d & u\bar s & u\bar c\\
d\bar u & d\bar d & d\bar s & d\bar c\\
s\bar u & s\bar d & s\bar s & s\bar c\\
c\bar u & c\bar d & c\bar s & c\bar c\\
			      \end{array}\right)\,,
\label{eq:P1}
\end{equation} 
which represents the quark-antiquark representation of the
pseudoscalar meson matrix. In the physical meson basis, including the standard $\eta$-$\eta'$ mixing \cite{Bramon:1992kr}, one has
\begin{eqnarray}
P=
\left(\begin{array}{cccc} 
              \frac{\pi^0}{\sqrt{2}}  + \frac{\eta}{\sqrt{3}}+\frac{\eta'}{\sqrt{6}}& \pi^+ & K^+ & \bar D^0\\
              \pi^-& -\frac{1}{\sqrt{2}}\pi^0 + \frac{\eta}{\sqrt{3}}+ \frac{\eta'}{\sqrt{6}}& K^0 & D^-\\
              K^-& \bar{K}^0 & -\frac{\eta}{\sqrt{3}}+ \frac{2\eta'}{\sqrt{6}} & D_s^-\\
	      D^0 & D^+ & D_s^+ & \eta_c
      \end{array}
\right)\,
\label{eq:P2}
\end{eqnarray}

Using this representation, the hadronized state of Eq.~\eqref{eq:Hhadrons} can be expressed as
\begin{align}
|H\rangle= &\frac{1}{\sqrt{2}} D^{*-}_s\bigg( 
(\frac{\pi^0}{\sqrt{2}}  + \frac{\eta}{\sqrt{3}}+\frac{\eta'}{\sqrt{6}}) c u d
-\pi^- cuu 
\nn \\
&+\pi^+ cdd- (-\frac{1}{\sqrt{2}}\pi^0 + \frac{\eta}{\sqrt{3}}+ \frac{\eta'}{\sqrt{6}}) cdu \nn \\
&+K^+ csd -K^0 csu + \bar D^0 ccd - D^- ccu
\bigg),
\label{eq:Hhadrons}
\end{align}
where, for clarity, the bra-ket notation has been omitted.

The only term in Eq.~\eqref{eq:Hhadrons} that overlaps with one of the channels responsible for generating the $P_{css}$ state from $VB$ in our model is the $D_s^{*-} K^+ csd$ component. In particular, it couples to $D_s^{*-}\Xi_c^{\prime 0}$, whose wave function is
 \begin{eqnarray}
\Xi'^0_c=\frac{1}{\sqrt{2}}(cds+csd)\chi_{MS}(23),
\label{eq:X0csym}
\end{eqnarray}
where $\chi_{MS}$ denotes the mixed-symmetric spin wave function under the exchange of quarks 2 and 3~\cite{close_book,Wang:2022aga}.

At first glance, one might expect a vanishing overlap between the initial and final baryon wave functions, since the initial $\Lambda_b^0$ carries a mixed-antisymmetric spin function $\chi_{MA}(23)$ for the two spectator quarks. However, the situation changes after the weak transition and hadronization. The weak decay produces the configuration $c(1)u(2)d(3)$ with quarks (2) and (3) in an antisymmetric spin state. Following hadronization, the system becomes $c(1)s(4)\bar s(5)u(2)d(3)$, and the $\Xi_c^{\prime 0}$ forms from the subset $c(1)s(4)d(3)$.
In the $^3P_0$ model, the created $q\bar q$ pair carries $L=1$ and $S=1$, ensuring vacuum quantum numbers. As a result, the spin correlation between the $s$ and $d$ quarks is not restricted to the antisymmetric $\chi_{MA}$ component, allowing for a nonzero overlap with the $\Xi_c^{\prime 0}$ wave function. In other words, the new $s$ quark is not forced to couple strictly antisymmetrically with the original $d$ quark.
While an explicit evaluation of this overlap within the $^3P_0$ model is, in principle, possible (see, e.g., Ref.~\cite{Dai:2018thd,Dai:2018vzz,Liang:2018rkl}), such an effort is not worthwhile here, since the overall production strength remains undetermined.

\begin{figure}[h]
\centering
\includegraphics[width=.95\linewidth]{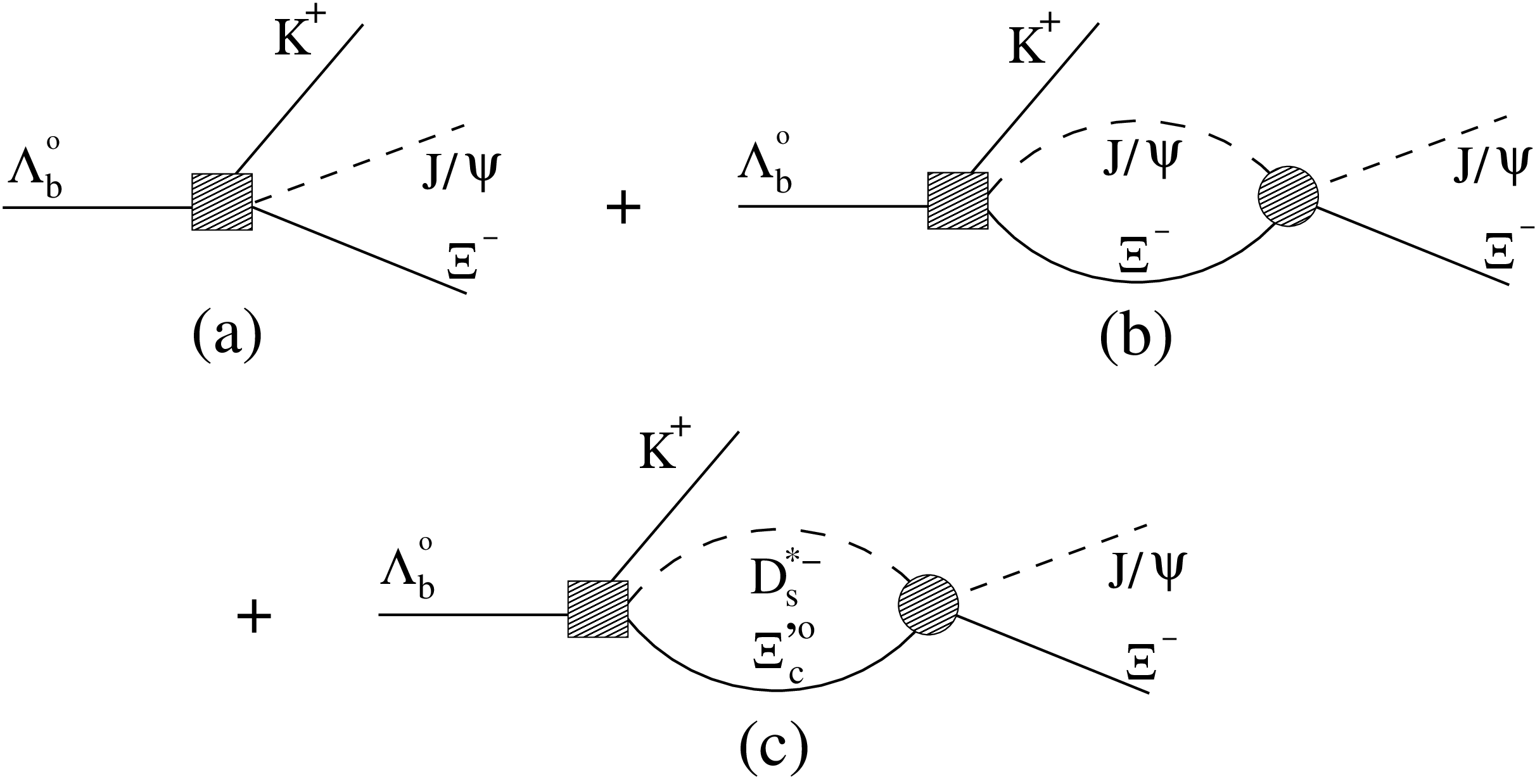}
\caption{Final state interaction of the meson-baryon pair. The square denotes the elementary production processes of Fig.~\ref{fig:diagram_tree_Lambda}, and the circle the coupled channel meson-baryon scattering amplitudes from Eq.~\eqref{eq:BS}.}
\label{fig:diag_tre_FSI_Lambdab}
\end{figure}

At this stage, the produced state is $D^{*-}_s \Xi_c^{\prime 0}K^+$. To obtain the final $J/\psi \Xi^- K^+$ state, and allow for the generation of the $P_{css}$, we need to implement the final state interaction (FSI) of the $D^{*-}_s\Xi_c^{\prime 0}$ pair leading to  $J/\psi \Xi^-$, as illustrated in Fig.~\ref{fig:diag_tre_FSI_Lambdab}(c).

In Fig.~\ref{fig:diag_tre_FSI_Lambdab}(c), the square represents the tree level $\Lambda_b\to D^{*-}_s \Xi_c^{\prime 0}K^+$ production, while the circle denotes the full coupled-channel vector-baryon scattering amplitude, $t_{12}$ from Eq.~\eqref{eq:BS}, which dynamically generates the $P_{css}$ state.

Taking the previous considerations into account, the decay amplitude for the external W-emission contribution to 
$\Lambda_b\to J/\psi \Xi^- K^+$, Fig.~\ref{fig:diag_tre_FSI_Lambdab}(c), reads

\begin{align}\label{eqn:fullamplitudeMee}
\mathcal{M}_{ee}(M_{\rm inv})={\cal C}G_{D^{*-}_s \Xi_c^{\prime 0}}(M_{\rm inv})\,t_{D^{*-}_s \Xi_c^{\prime 0},J/\psi \Xi^-}(M_{\rm inv}) \,,
\end{align}
where ${\cal C}$ accounts for the tree level amplitude  of the external W-emission process  (Fig.~\ref{fig:diagram_tree_Lambda}(a)) producing $D^{*-}_s \Xi_c^{\prime 0} K^+$, incorporating both the weak transition and hadronization dynamics.

For the internal W-emission mechanism, shown in Fig.~\ref{fig:diagram_tree_Lambda}(b), a similar procedure can be followed. In this case,  the $c\bar c$ pair produces the $J/\psi$ directly, and the hadronization requires creating an $s\bar s$ pair to form the kaon. Consequently, the relevant contribution corresponds to the $K^+$ term in Eq.\eqref{eq:Hhadrons}, but with the $s$ quark replacing the original $c$, yielding a $K^+ ssd$ term that overlaps with the $\Xi^-$ wave function.

\begin{eqnarray}
\Xi^-=\frac{1}{\sqrt{2}} \left( \phi_{MS}\chi_{MS}+\phi_{MA}\chi_{MA}\right),
\label{eq:Ximenoswf}
\end{eqnarray}
with $\phi$ standing for the flavor wave function.

In the internal W-emission mechanism, the $J/\psi \Xi^- K^+$ final state is generated already at tree level (Fig.~\ref{fig:diag_tre_FSI_Lambdab}(a)) and can subsequently undergo final-state interactions, dynamically producing the $P_{css}$ state (Fig.~\ref{fig:diag_tre_FSI_Lambdab}(b)). The corresponding amplitude can be expressed as
\begin{align}\label{eqn:fullamplitudeMie}
\mathcal{M}_{ie}(M_{\rm inv})={\cal C}\frac{1}{N_c}\left( 1+G_{J/\psi \Xi^-}(M_{\rm inv})\,t_{J/\psi \Xi^-,J/\psi \Xi^-}(M_{\rm inv}) \right)\,,
\end{align}
The factor $1/N_c$ comes from the fact that the external W-emission is amplified by the number of colors, $N_c = 3$, due to the sum over quark colors in the hadronization, unlike the internal emission where the final meson color is fixed. On the other hand, the relative sign between these two contributions is ambiguous and could, in principle, affect their interference. However, the decay $\Lambda_b \to J/\psi \Xi^- K^+$ is  dominated by the external-emission process, making this ambiguity practically irrelevant.

At this point, it is worth commenting on the spin structure implicitly contained in the factor ${\cal C}$ in Eqs.~\eqref{eqn:fullamplitudeMee} and \eqref{eqn:fullamplitudeMie}.
The  polarization four-vector $\epsilon^\mu$ must be contracted with another four vector. Following Ref.~\cite{Lyu:2025rsq}, for processes like $B^+\rightarrow D^{*-}D_s^+\pi ^+$, $\epsilon^\mu$  should be contracted with the initial decaying baryon four-momentum  $P^\mu$,  which in our case  gives a vertex

\begin{align}\label{eqn:spinvertex}
{\cal C}={\cal C'}P^\mu \epsilon_\mu
\end{align}
with ${\cal C'}$ a constant.
Averaging over initial $\Lambda_b$ polarization and summing over final vector polarizations yields a factor in the squared amplitude proportional to
\begin{align}\label{eqn:P2}
-P^2+\frac{(P\cdot q)^2}{M^2_{J/\psi}},
\end{align}
with $q$ the $J/\psi$ momentum.

 Finally, the $J/\psi \Xi^-$ invariant mass distribution  can be expressed as:
\begin{align}\label{eqn:dGammadM}
\frac{d\Gamma}{dM_{\rm inv}}(M_{\rm inv})
=\frac{1}{(2\pi)^3}\frac{M_{\Xi^-}}{M_{\Lambda_b}}p_{K} \tilde p_{J/\psi}\,F\,\left|\mathcal{M}(M_{\rm inv})\right|^2\,,
\end{align}
where $p_{K}$ and $\tilde p_{J/\psi}$ denote the $K^+$ momentum in the $\Lambda_b$ rest-frame, and the $J/\psi$ momentum in the final $J/\psi \Xi^-$ center-of-mass frame, respectively,
\begin{align}
& p_k=\frac{\lambda^{1/2}\left(M_{\Lambda_ b}^2 M_\text{inv}^2, M_{K^+}^2\right)}{2 M_{\Lambda_b}} \\
& \tilde{p}_{J/ \psi}=\frac{\lambda^{1 / 2}\left(M_\text{inv}^2, M_{J/ \psi}^2, M_{\Xi^-}^2\right)}{2 M_\text{inv}}, 
\end{align}
with $\lambda$ being the usual K\"all\'en function.

In Eq.~\eqref{eqn:dGammadM}, the total amplitude is given by $\mathcal{M}=\mathcal{M}_{ee}+\mathcal{M}_{ie}$, with $\mathcal{M}_{ee}$ and $\mathcal{M}_{ie}$ given in Eqs.~\eqref{eqn:fullamplitudeMee} and \eqref{eqn:fullamplitudeMie}. The factor $F$, coming from Eq.~\eqref{eqn:P2}, is
\begin{align}\label{eqn:Ffactor}
F=\frac{1}{M_{J/\psi}^2}\left[\tilde P^{0^2} \tilde q^{0^2}
+\frac{1}{3}\left(|\tilde P||\tilde q|\right)^2\right]-M_{\Lambda_b}^2
\end{align}
where
\begin{align}
&\tilde{P}^0=\frac{1}{2 M_\text{inv}}\left(M_{\Lambda_b}^2+M_\text{inv}^2-M_{K^+}^2\right) \nn \\ &\tilde{q}^0=\frac{M_\text{inv}^2+M_{J/ \psi}^2-M_{\Xi^-}^2}{2 M_\text{inv}} \nn \\
& \tilde{P}=\frac{\lambda^{1 / 2}\left(M_{\Lambda_b}^2, M_\text{inv}^2, M_{K^+}^2\right)}{2 M_\text{inv}} \nn \\
& \tilde{q}=\tilde{p}_{J/ \psi}.
\end{align}

For the evaluation of the $\Xi_b\to J/\psi \Xi^- \pi^+$, we must consider that
now the wave function of the initial $\Xi^0_b$ is \cite{Capstick:1986ter,Roberts:2007ni,Wang:2022aga}
\begin{align*}
|\Xi_b^0\rangle=|b(us-su)\rangle\,\chi_{MA},
\end{align*}
The external and internal W-emission mechanisms proceed analogously to those in Fig.~\ref{fig:diagram_tree_Lambda}, with the $d$ quark in the lower line replaced by an $s$ quark.
The main difference from $\Lambda_b\to J/\psi \Xi^- K^+$ lies in the hadronization, which now requires generating a $d\bar d$ pair to produce the final $\pi^+$. This leads to a formalism where the figures, amplitudes and decay width expressions remain the same with $\Lambda_b$ replaced by $\Xi_b$ and $K^+$ replaced by $\pi^+$.


\section{Results}

\begin{figure}[h]
     \centering
     \subfigure[]{\includegraphics[width=.95\linewidth]{Minv_L.eps}} \\
     \subfigure[]{\includegraphics[width=.95\linewidth]{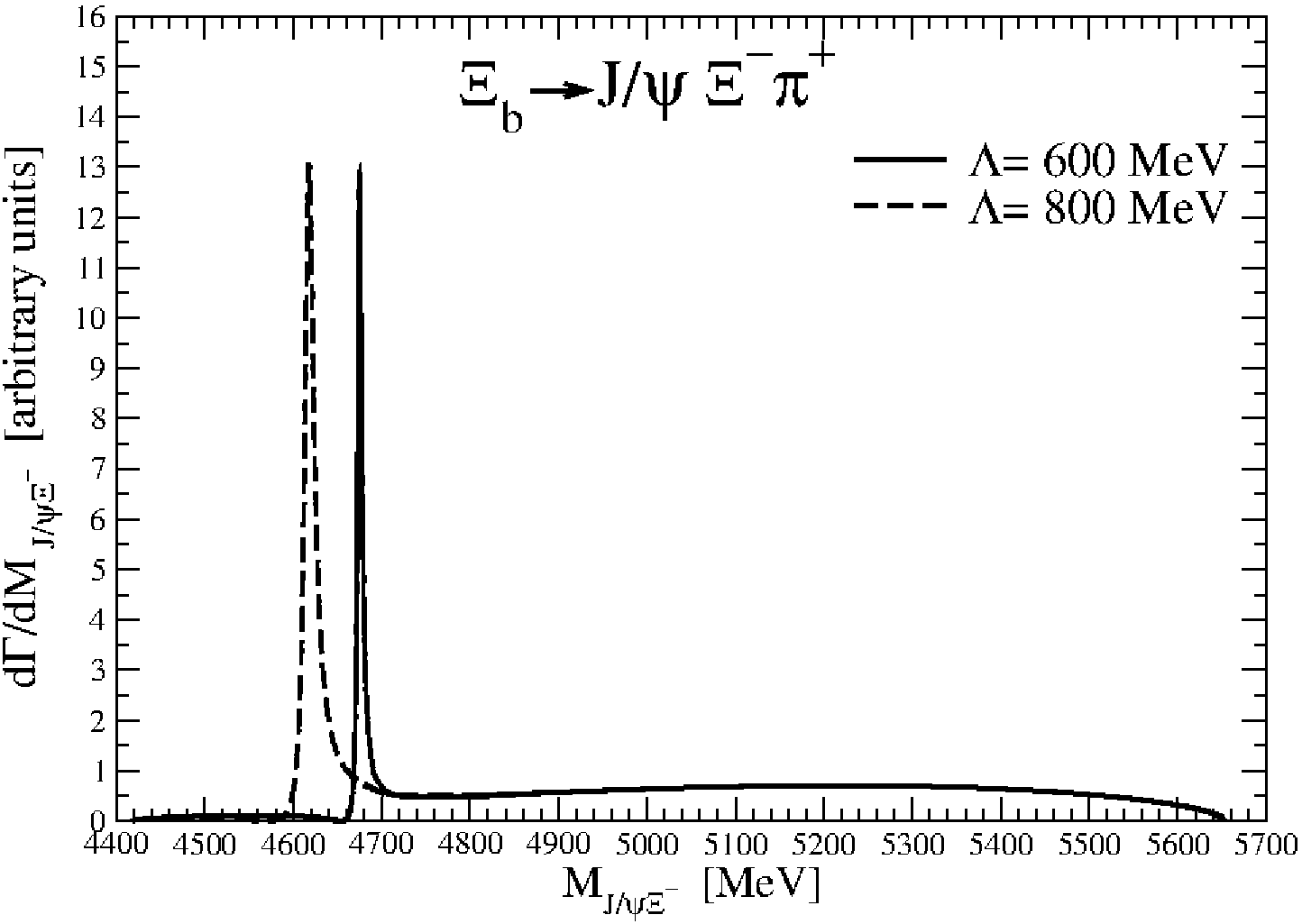}} \\
    \caption{The $J/\psi \Xi^-$  invariant mass distributions for the $\Lambda_b\to J/\psi \Xi^- K^+$ and $\Xi_b\to J/\psi \Xi^- \pi^+$ decays, for two different values of the meson-baryon loop regularization cutoff, $\Lambda$.
}
\label{fig:results1}
\end{figure}

In Fig.~\ref{fig:results1} we show the $J/\psi \Xi^-$ invariant mass distributions for both decays considered in this work, evaluated with the two values of the loop regularization cutoff used in Ref.\cite{Roca:2024nsi}, $\Lambda = 600$ and $800$~MeV. The difference between the two curves provides an estimate of the dominant source of theoretical uncertainty. In the results shown in the remainder of the paper we use $\Lambda = 600$~MeV as a reference, keeping in mind an uncertainty of the magnitude illustrated in Fig.~\ref{fig:results1}.

The spectra show pronounced narrow peaks associated to the poles generated in the coupled-channel interaction for the $I(J^P)=\tfrac{1}{2}(\tfrac{1}{2}^-,\tfrac{3}{2}^-)$ in the $VB$ channel: $M = 4675$ MeV and $\Gamma = 10$~MeV  $\Lambda=600$ MeV; and  $M = 4617$~MeV and $\Gamma = 12$~MeV for $\Lambda=800$ MeV.
It should be emphasized that the highly non-linear dynamics involved in the unitarized coupled-channel amplitudes in Eq.~\eqref{eq:BS} provides not only the pole positions but also the full, nontrivial, line shapes of the vector-baryon amplitudes, which are not necessarily of Breit-Wigner type.

While these results might suggest that the predicted resonances could, in principle, be easily identified in dedicated experimental measurements of the $J/\psi\Xi^-$ invariant mass distribution, the situation is more involved. The very narrow widths expected for these $P_{css}$ states, together with potential background contributions not yet accounted for, make an unambiguous observation less straightforward, as discussed below.

\begin{figure}[h]
     \centering
     \subfigure[]{\includegraphics[width=.95\linewidth]{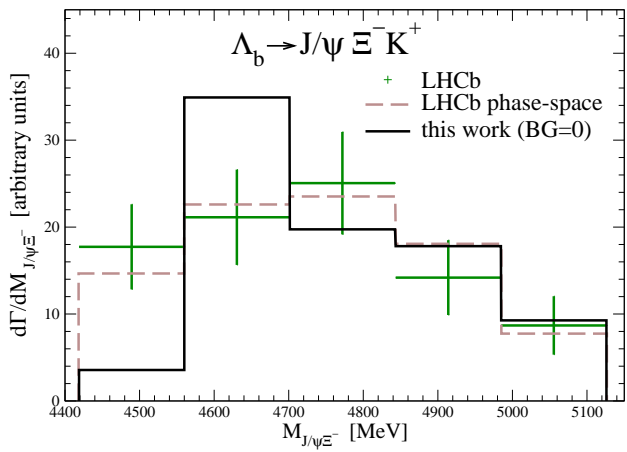}} \\
     \subfigure[]{\includegraphics[width=.95\linewidth]{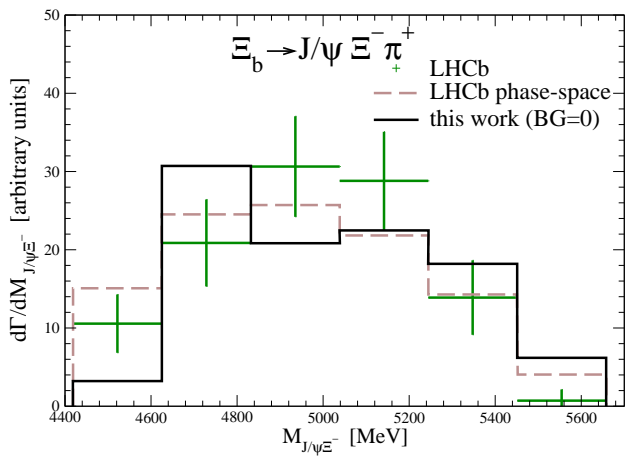}} \\
    \caption{Experimental $J/\psi \Xi^-$ mass distribution (crosses) and experimental simulated phase-space (dashed histograms) from LHCb  \cite{LHCb:2025lhk}. The solid lines show our theoretical results averaged over the experimental bin width.   
}
\label{fig:results2}
\end{figure}

\begin{figure}[h]
     \centering
     \subfigure[]{\includegraphics[width=.95\linewidth]{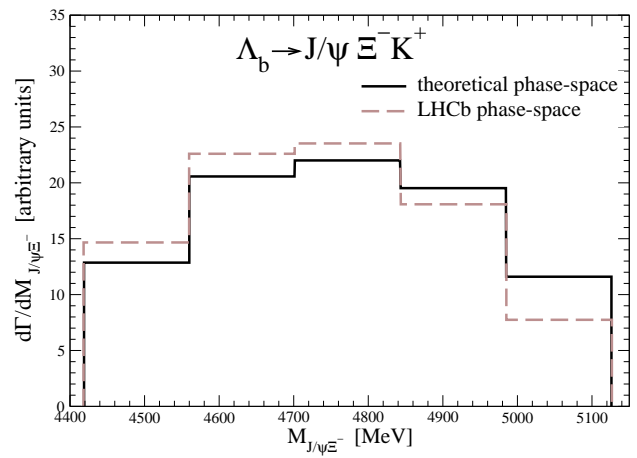}} \\
     \subfigure[]{\includegraphics[width=.95\linewidth]{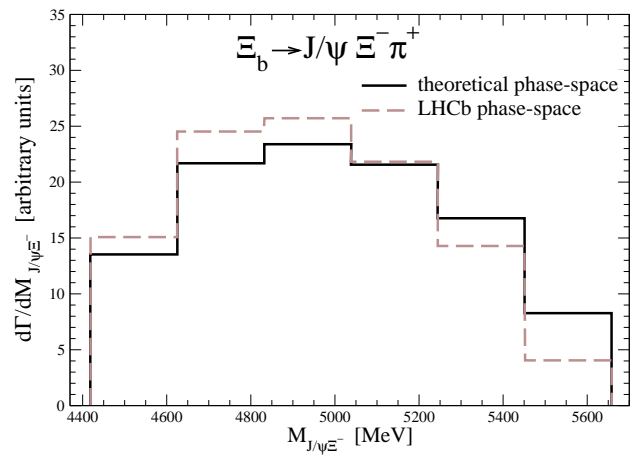}} \\
    \caption{Experimental simulated phase-space from LHCb setup (dashed lines) \cite{LHCb:2025lhk} and theoretical phase-space averaged over the same bins. 
}
\label{fig:results3}
\end{figure}

At this point we can take advantage of  recent experimental measurements by the LHCb Collaboration \cite{LHCb:2025lhk}. The corresponding experimental $J/\psi \Xi^-$ spectra are shown in Fig.~\ref{fig:results2}. 
The crosses represent the measured data with errors, and the dashed histograms correspond to the simulated experimental phase-space. The measured mass resolution is $141$~MeV for $\Lambda_b\to J/\psi \Xi^-K^+$ and $206$~MeV for $\Xi_b\to J/\psi\Xi^-\pi^+$. The solid lines show our theoretical distributions averaged over bins of the same width as in the experiment. 
Since the global normalization is arbitrary, we have rescaled our results so that the total area of each theoretical distribution matches the corresponding experimental one.
 Another point to be considered is that the simulated experimental phase-space is affected by detector acceptance and therefore does not perfectly match a purely theoretical phase-space distribution binned with the same resolution, as illustrated in Fig.~\ref{fig:results3}. To enable a meaningful comparison between theory and experiment, we have modified our theoretical results by multiplying each bin by the ratio of the experimental phase-space to the theoretical one.

In the experimental data in Fig.~\ref{fig:results2}, no disagreement between the measurement and the simulated phase-space is observed  within  uncertainties, and no clear signal of the possible resonant states is apparent. 
This is not surprising, given that the widths of the $P_{css}$ states of interest are an order of magnitude smaller than the experimental bin size, so any potential peaks would be effectively averaged within a single large bin.

\begin{figure}[h]
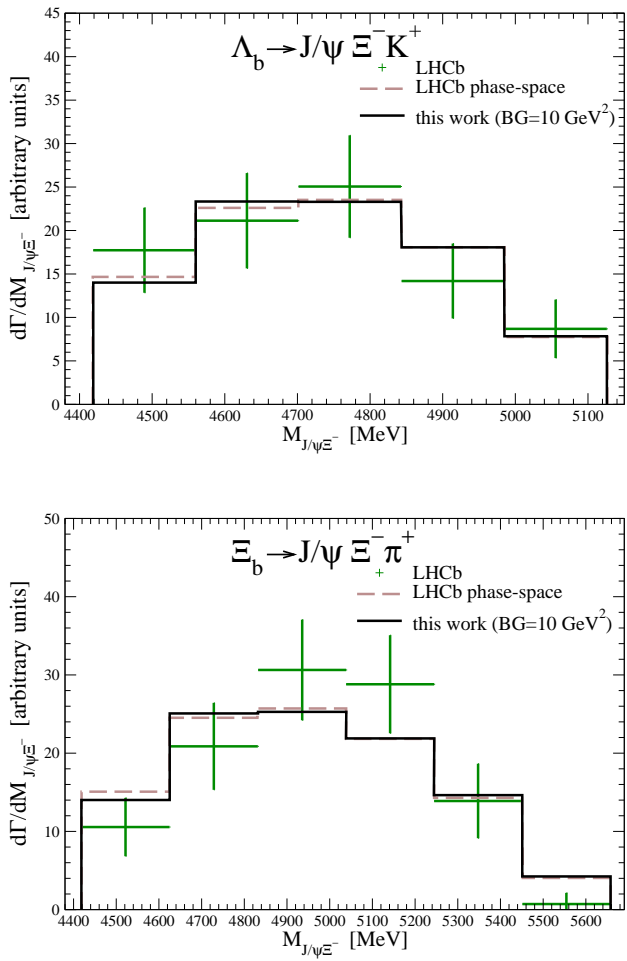

     \centering
    \subfigure[]{\includegraphics[width=.95\linewidth]{Minv_exp_th_L_BG_hist.eps}} \\
     \subfigure[]{\includegraphics[width=.95\linewidth]{Minv_exp_th_X_BG_hist.eps}} \\
    \caption{ Same as Fig.~\ref{fig:results3} but for a background of $10\,GeV^2$.
}
\label{fig:results4}
\end{figure}

On the other hand, the theoretical binned distribution, while showing an enhancement in the resonant region, does not appear fully consistent with the experimental data. However, to make a meaningful comparison, one should consider that additional background contributions may be present. Within our model, some nontrivial background structure naturally emerges beyond the resonances due to the unitarized coupled-channel dynamics. The tree-level internal W-emission process, Fig.~\ref{fig:diag_tre_FSI_Lambdab}(a), also provides a background contribution. Nevertheless, other sources of background may be present in the experiment.
Since the goal of this discussion is to assess the feasibility of observing the $P_{css}$ resonances in the experimental distributions, we introduce an incoherent background term, $BG$, into Eq.~\eqref{eqn:dGammadM} as:
\begin{align}\label{eqn:dGammadM_BG}
\frac{d\Gamma}{dM_{\rm inv}}(M_{\rm inv})
=\frac{1}{(2\pi)^3}\frac{M_{\Xi^-}}{M_{\Lambda_b}}p_{K} \tilde p_{J/\psi}\,\left(F\,\left|\mathcal{M}(M_{\rm inv})\right|^2\,+BG\right),
\end{align}
As an illustrative example, using a background of $BG = 10~\text{GeV}^2$, Fig.~\ref{fig:results4} shows a distribution compatible with the experimental data, though with little sign of the resonant states. Indeed, it is barely distinguishable from phase-space.

\begin{figure}[h]
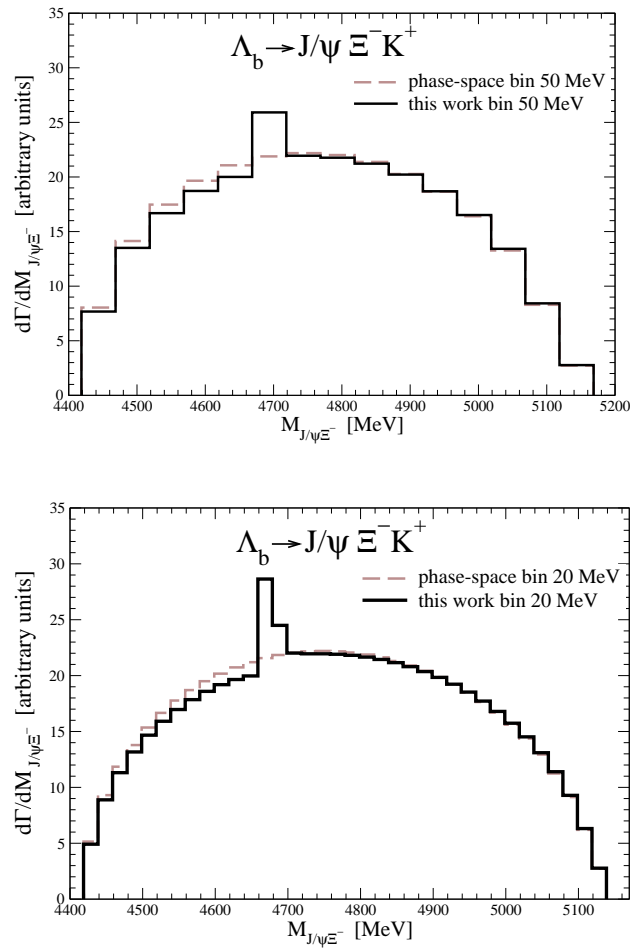

     \centering
     \subfigure[]{\includegraphics[width=.95\linewidth]{Minv_bin50_BG_hist.eps}} \\
     \subfigure[]{\includegraphics[width=.95\linewidth]{Minv_bin20_BG_hist.eps}} \\
     \caption{ Theoretical predictions (solid lines) for the $J/\psi \Xi^-$ spectrum, in $\Lambda_b\to J/\psi \Xi^- K^+$ decay and theoretical phase-space (dashed lines) for to different widths of the bins, $50~MeV$ in the upper panel and $20~MeV$ in the lower one,  for a background of $10\,GeV^2$.
}
\label{fig:results5}
\end{figure}

To elaborate on the experimental mass resolution required to resolve the resonances, Fig.~\ref{fig:results5} shows the theoretical distribution
for $\Lambda_b\to J/\psi \Xi^- K^+$,  including a background of  $BG=10\,GeV^2$, averaged over bin widths of 50~MeV (upper panel) and 20~MeV (lower panel). The solid lines represent the full model prediction, while the dashed lines correspond to the binned phase-space. The  $\Xi_b\to J/\psi \Xi^- \pi^+$ case leads to qualitatively similar conclusions and is not shown.

The difference between the full model and the phase-space curves provides an indication of the experimental precision needed to resolve these narrow resonances.
Although we include a background term mainly to match the overall normalization of the experimental spectra, our underlying mechanism is expected to account for a significant fraction of the observed strength. Consequently, the peaks shown in Fig.~\ref{fig:results5} should be regarded as a conservative estimate, and the actual resonant signals could be more pronounced if the true background level is smaller than assumed.

\section{Conclusions}

In this work, we have investigated theoretically the possibility of observing a hidden-charm, double-strangeness pentaquark with $I(J^P)=\tfrac{1}{2}(\tfrac{1}{2}^-,\tfrac{3}{2}^-)$ in the $J/\psi \Xi^-$ invariant-mass spectrum of the decays $\Lambda_b\to J/\psi \Xi^- K^+$ and $\Xi_b\to J/\psi \Xi^- \pi^+$.
These $P_{css}$ states, with flavor content $\bar c c s s n$, were predicted in earlier theoretical studies based on unitarized coupled-channel dynamics. In this framework, which closely follows the chiral unitary approach, meson-baryon scattering amplitudes are unitarized starting from tree-level vector-baryon potentials derived from $t$-channel vector-meson exchange. The corresponding Lagrangians are constructed from extensions of the local hidden-gauge formalism to the heavy-quark sector. Within this framework, pentaquarks emerge dynamically as poles on unphysical Riemann sheets, without being introduced as explicit degrees of freedom.

To apply this formalism to the $\Lambda_b\to J/\psi \Xi^- K^+$ and $\Xi_b\to J/\psi \Xi^- \pi^+$ decays, we first identify the dominant weak decay mechanisms at tree level, involving both internal and external $W$-emission topologies. After the weak transition, hadronization is implemented following the $^3P_0$ model, generating $q\bar q$ pairs and allowing the production of $J/\psi \Xi^-$ and $D_s^{*-} \Xi_c^{\prime 0}$ pairs, which are the main baryon-vector channels generating the pentaquark poles. These channels are then incorporated in a coupled-channel vector-baryon interaction to dynamically produce the pentaquark resonance.

After folding the predicted $J/\psi \Xi^-$ invariant-mass distributions with the current LHCb bin resolutions ($\gtrsim100$-200~MeV), the narrow resonant signal is strongly diluted, explaining the absence of a visible structure in present data. We also show how the predicted signal becomes visible with improved mass resolution. These results should motivate future upgrades in experimental resolution  aimed at probing such hadronic states.

\section*{ACKNOWLEDGEMENTS}

This
work is partly supported by the Spanish Ministerio
de Economia y Competitividad (MINECO) and European FEDER funds under
Contracts No. FIS2017-84038-
C2-1-P B, PID2020-112777GB-I00, and by Generalitat
Valenciana under contract PROMETEO/2020/023. This
project has received funding from the European Union
Horizon 2020 research and innovation programme under
the program H2020-INFRAIA-2018-1, grant agreement
No. 824093 of the STRONG-2020 project.
This work is partly supported by the Science and Technology Facilities Council (UK).
This work is partly supported by the National Natural Science Foundation of China under Grants No. 12405089 and No. 12247108 and the China Postdoctoral Science Foundation under Grant No. 2022M720360 and No. 2022M720359.


\end{document}